\pdfoutput=1
\documentclass{iopjournal}
\usepackage{cmap}
\usepackage{xcolor}
\usepackage[utf8]{inputenc}
\usepackage[T1]{fontenc}
\usepackage[numbers,sort&compress]{natbib}
\usepackage{graphicx,makecell,subcaption,amsmath,booktabs}

\begin{document}


\title{Quasinormal spectra of higher dimensional regular black holes in theories with infinite curvature corrections}

\author{Juan Pablo Arbelaez$^1$\orcid{0009-0006-3313-4584}}

\affil{$^1${Centro de Matemática, Computação e Cognição (CMCC), Universidade Federal do ABC (UFABC),
Rua Abolição, CEP: 09210-180, Santo André, SP, Brazil}}

\email{juan.arbelaez@ufabc.edu.br}


\begin{abstract}
We investigate the quasinormal modes of massless scalar and electromagnetic test fields propagating on several families of higher-dimensional regular black holes arising in gravitational theories with an infinite tower of higher-curvature corrections to Einstein gravity. Using the Wentzel--Kramers--Brillouin (WKB) method supplemented by Padé approximants, we compute the fundamental modes and derive compact analytic expressions in the eikonal regime. For all models considered, positive higher-curvature coupling reduces both the oscillation frequencies and damping rates relative to the corresponding singular general-relativistic limit. Since our analysis is restricted to test-field perturbations on fixed regular black-hole backgrounds, these results should be interpreted as geometric and phenomenological signatures of the background geometry rather than as gravitational perturbations of the full higher-curvature theory. Our results provide further insight into how quantum-gravity-inspired corrections and regularity conditions may affect black-hole ringdown signals.
\end{abstract}

\section{Introduction}

Quasinormal modes (QNMs) of black holes describe the characteristic oscillations of spacetime geometry in response to perturbations \cite{Kokkotas:1999bd,Berti:2009kk,Nollert:1999ji,Konoplya:2011qq,Bolokhov:2025uxz}. These modes appear as damped oscillations with complex frequencies, whose real part governs the oscillation frequency while the imaginary part determines the damping time. Since QNMs depend only on the macroscopic parameters of the black hole, they provide a direct fingerprint of the underlying geometry. With the advent of gravitational-wave astronomy, in particular through the observations of the LIGO and Virgo Collaborations \cite{LIGOScientific:2016aoc,LIGOScientific:2017vwq,LIGOScientific:2020zkf}, and more recently KAGRA \cite{KAGRA:2022qtq}, QNMs have become an observational reality. The initial phase of a signal from black-hole merger is dominated by quasinormal ringing, and thus offers a unique opportunity to test gravity in the strong-field regime. Precise measurements of QNM spectra not only allow for independent determinations of black-hole mass and spin, but also enable stringent tests of the no-hair theorem, searches for exotic compact objects, and potential deviations arising from quantum gravity or higher-curvature corrections \cite{Konoplya:2022pbc,Horowitz:2023xyl}.

Despite the successes of classical general relativity, black-hole solutions, such as the Schwarzschild or Kerr metrics, suffer from the presence of a central curvature singularity, where spacetime invariants diverge and the classical description breaks down \cite{Christodoulou:1991yfa}. This pathology signals the incompleteness of general relativity as a fundamental theory, and it is widely expected that quantum gravity effects should resolve or regularize such singularities. Various approaches, ranging from loop quantum gravity and string-inspired models to effective modifications of Einstein's equations \cite{Reuter:2019byg,Bonanno:2001hi,Bonanno:2002zb,Rubano:2004jq,Koch:2013owa,Pawlowski:2018swz,Ishibashi:2021kmf}, suggest that the classical singularity can be replaced by a regular core, thereby giving rise to such regular black holes. These spacetimes preserve the essential features of black holes, such as the event horizon, while avoiding divergences at the center.

Recently, regular black holes were constructed in \cite{Bueno:2024dgm}, arising purely from gravity in higher-dimensional spacetimes through the inclusion of an infinite tower of higher-curvature corrections. The infinite tower of corrections can be interpreted as a resummed effective description or phenomenological completion of the high-curvature regime, rather than as a fundamental ultraviolet completion of gravity. The convergence of the series and the positivity conditions imposed on the couplings are essential because they make the resummed function $h(\psi)$ well defined in the relevant domain and allow the near-origin geometry to approach a finite-curvature core. In four-dimensional spacetimes, upon spherical reduction, these theories map to a particular subclass of Horndeski scalar–tensor models~\cite{Bueno:2025zaj}. Appropriate choices of the coupling constants lead to several families of regular black holes \cite{Bueno:2024zsx,Bueno:2024eig,Bueno:2025gjg}, including $D$-dimensional generalizations of well-known regular metrics, such as the Hayward and Bardeen-type solutions.

Higher-dimensional spacetimes arise naturally in several approaches to quantum gravity, where additional spatial dimensions are introduced as a fundamental ingredient of the underlying theory. In particular, frameworks such as string theory suggest that, at length scales approaching the Planck scale, spacetime may effectively exhibit more than four dimensions \cite{Boulware:1985wk}. From the perspective of effective field theory, General Relativity can be understood as the low-energy limit of a more fundamental description of gravity, where higher-curvature corrections appear as subleading contributions that become increasingly relevant in high-curvature regimes. Black holes, especially those with small horizon radius, provide a natural setting in which such effects may become important. In this context, higher-dimensional black hole solutions offer a useful theoretical laboratory to explore deviations from General Relativity and the role of regularity in strong-gravity regimes \cite{Rychkov:2004sf}.

A particularly compelling motivation for studying higher-dimensional gravity comes from models with large or warped extra dimensions. In the Arkani-Hamed–Dimopoulos–Dvali model \cite{Arkani-Hamed:1998jmv,Arkani-Hamed:1998sfv}, the presence of large, compact extra dimensions lowers the fundamental Planck scale to values potentially accessible at high-energy colliders, thereby opening the possibility of producing microscopic black holes in particle collisions. These objects would have horizon radii comparable to the size of the extra dimensions and would evaporate rapidly via Hawking radiation, providing distinctive signatures that probe the underlying higher-dimensional geometry. In contrast, the Randall-Sundrum models \cite{Randall:1999ee,Randall:1999vf}, which involve a non-factorizable warped geometry, localize gravity near a four-dimensional brane while still allowing for strong gravitational effects at TeV scales. In such settings, mini black holes may also form, though their properties are influenced by the warped extra dimension and brane tension \cite{Giddings:2002kt}. In both scenarios, the study of these small black holes highlights the importance of higher-dimensional corrections and suggests that deviations from classical General Relativity could become observable in regimes where curvature is large but still potentially accessible to experiment or phenomenology \cite{Kanti:2008eq}.

The quasinormal modes of these latter models were analyzed in \cite{Konoplya:2024hfg}. It was shown that higher overtones are more sensitive to the higher-curvature couplings and can exhibit features such as vanishing real parts, corresponding to purely decaying modes. From the observational perspective, however, the fundamental (dominant) mode is of primary importance, as it governs the late-time ringdown signal and thus leaves the strongest imprint in the detected waveform. Higher overtones decay much more rapidly, making them increasingly difficult to resolve with current gravitational-wave detectors. Consequently, the dominant mode provides the most robust and accessible probe of black hole parameters and potential deviations from general relativity. In the present work, we extend previous analyses in two directions:
\begin{itemize}
  \item We examine all the proposed models emerging from the higher curvature expansion and include a new one that has not been analyzed before, in addition to Hayward- and Bardeen-type black holes.

  \item We systematically investigate how the degree and nature of regularity of the black hole influence both the real and imaginary parts of the QNMs. In particular, we look for general trends among the models, and aim to understand how regularity modifies oscillation frequencies and damping times, focusing on the fundamental modes.
\end{itemize}
It is important to emphasize that higher-dimensional regular black holes are not directly motivated by astrophysical observations. Rather, they arise within higher-curvature extensions of gravity, where an effective-field-theory framework naturally incorporates an infinite tower of curvature corrections beyond the Einstein–Hilbert action. In this sense, these solutions provide a theoretically consistent setting to investigate how regularity emerges and how it affects black-hole dynamics. By considering several distinct families of regular black hole solutions generated within this framework, rather than focusing on a single metric, we aim to identify robust and model-independent features of regular black holes and to contrast them with the quasinormal mode spectra of singular black holes in General Relativity.

This paper is organized as follows. Sec.~\ref{sec:basic} is devoted to the construction of the regular black-hole solutions within the theory with an infinite tower of higher-order curvature corrections, and their perturbation equations. In Sec.~\ref{sec:WKB} we review the WKB method used for calculating the quasinormal modes. In Sec.~\ref{sec:qnms} we present and discuss the obtained results. Finally, Sec.~\ref{sec:conclusions} offers concluding remarks and possible directions for future work.

\section{Basic equations}\label{sec:basic}

Following \cite{Bueno:2019ycr}, we consider gravitational theory, given by the action
\begin{equation}
    I_{QT} = \frac{1}{16\pi G}\int{d^D x\sqrt{|g|}\left[R+\sum^{n_{max}}_{n=2}{\alpha_n \mathcal{Z}_n}\right]},
    \label{eq:action}
\end{equation}
where \(G\) is the \(D\)-dimensional Newton constant and the quantities \(\mathcal{Z}_n\) denote quasi-topological curvature invariants of order \(n\). The explicit expressions for the first five densities \(\mathcal{Z}_n\) can be found in Appendix A of Ref.~\cite{Bueno:2024dgm}, while a general expression valid for arbitrary order \(n\) is given in Eq.~(49) of Ref.~\cite{Bueno:2019ycr}. The coefficients \(\alpha_n\) are arbitrary constant coupling parameters associated with higher-curvature corrections, which set the characteristic scale of the theory and have dimensions of length$^{2(n-1)}$. The static spherically symmetric solution is given by the following line element \cite{Bueno:2024dgm}:
\begin{equation}
    ds^2 = -N(r)^2 f(r)dt^2 + \frac{dr^2}{f(r)} + r^2d\Omega^2_{D-2},
    \label{eq:metric}
\end{equation}
where \(\Omega^2_{D-2}\) denotes the line element of the unit
\((D-2)\)-sphere, and with two unknown functions $N(r)$ and $f(r)$. Varying the reduced action with respect to these functions, we obtain
\begin{equation}
    \frac{dN}{dr}=0,\quad \frac{d}{dr}\left[r^{D-1}h(\psi)\right]=0,
\end{equation}
the first equation implies that \(N(r)\) is constant, and we fix \(N(r)=1\) by normalizing the time coordinate at infinity; the second equation can be integrated to yield
\begin{equation}
    h(\psi)=\frac{m}{r^{D-1}},
\label{eq:f(r)}
\end{equation}
here $m$ is an integration constant proportional to the ADM mass of the solution and where
\begin{equation}
    \label{eq:hpsi}
   h(\psi) \equiv \psi + \sum_{n=2}^{n_{\max}} \alpha_{n}\psi^{n},
   \qquad
   \psi\equiv\frac{1-f(r)}{r^{2}},
\end{equation}
which determines the metric function \(f(r)\).

As mentioned in Ref.~\cite{Bueno:2024dgm}, the conditions that must be imposed on the alpha coupling term are
\begin{equation}
    \alpha_n \geq 0 \quad\forall\;n\quad;\quad \limsup_{n\to\infty}{(\alpha_n)^{1/n}} = \frac{1}{C}>0,
\end{equation}
and must be fulfilled for the series to converge uniformly on every compact subset of the convergence disk. These mathematical restrictions have a direct physical role: they prevent alternating or non-convergent higher-curvature contributions from producing branch ambiguities in $h(\psi)$, and they ensure that the infinitely many terms can collectively modify the $r\to0$ behavior instead of leaving the highest finite-order correction dominant. In \cite{Bueno:2024dgm} the authors present several examples of this quasi-topological $\alpha$ coupling that give rise to regular black hole solutions when a tower of infinite terms is taken into account. Some of the elements presented there are well known in the literature and will be used for comparative purposes throughout this study.

Working with an infinite sum of higher-curvature correction terms is not arbitrary. For a finite number of terms, $n_{\rm max}=N$, analyzing the deep interior ($r\to 0$) shows that the highest-order term dominates:
\begin{equation}
    f(r)\approx 1-\left(\frac{m}{\alpha_{N}}\right)^{1/N} r^{\,2-(D-1)/N}.
\end{equation}

This situation presents several problems. First, the Kretschmann scalar diverges at $r=0$, so the singularity persists. Second, the lower-order terms, which could partially regulate the interior, are effectively suppressed.

These issues motivate considering the limit $N\rightarrow\infty$, which yields
\begin{equation}
    f(r)\sim 1-\Lambda\, r^{2}, \qquad \Lambda \equiv \lim_{N\rightarrow\infty}\left(\frac{m}{\alpha_{N}}\right)^{1/N}.
\end{equation}

In this limit, the spacetime near the origin becomes (anti-)de Sitter, with a finite curvature scalar. Physically, this shows that the singularity can be smoothed out by the cumulative effect of infinitely many higher-curvature terms. A general solution can then be computed that is valid across all regimes. It should be noted, however, that this mechanism relies critically on the existence of an infinite series of terms and the convergence of the corresponding limit, which must be justified within a consistent theoretical framework. Various choices of the coupling constants lead to regular black-hole configurations when the infinite tower of higher-curvature corrections is taken into account. In this paper, these choices are treated as controlled resummed models, useful for identifying robust consequences of regularity, rather than as unique predictions of a fundamental theory. Particular models are listed in table~\ref{tab:solutions} with additional properties listed in table~\ref{tab:ext-tab-sol}.


\begin{table}[t]
\centering
\caption{Summary of the considered regular black hole models:
Configurations $(a)$--$(e)$ were proposed in \cite{Bueno:2024dgm}.
The model $(f)$ is studied for the first time.}
\label{tab:solutions}
\setlength{\tabcolsep}{4pt}
\renewcommand{\arraystretch}{1.3}
\begin{tabular}{c c c c}
\hline
Label & $\alpha_n$ & $h(\psi)$ & $f(r)$ \\
\hline

$(a)$
&
$\displaystyle \alpha^{n-1}$
&
$\displaystyle \frac{\psi}{1-\alpha\psi}$
&
$\displaystyle 1-\frac{m r^2}{r^{D-1}+\alpha m}$ \\

$(b)$
&
$\displaystyle \frac{(1-(-1)^n)\Gamma\!\left(\frac{n}{2}\right)\alpha^{n-1}}
{2\sqrt{\pi}\Gamma\!\left(\frac{n+1}{2}\right)}$
&
$\displaystyle \frac{\psi}{\sqrt{1-\alpha^{2}\psi^{2}}}$
&
$\displaystyle 1-\frac{m r^2}{\sqrt{r^{2(D-1)}+\alpha^2 m^2}}$ \\

$(c)$
&
$\displaystyle \frac{\alpha^{n-1}}{n}$
&
$\displaystyle -\frac{\log(1-\alpha\psi)}{\alpha}$
&
$\displaystyle 1-\frac{r^2}{\alpha}
\left(1-e^{-\alpha m/r^{D-1}}\right)$ \\

$(d)$
&
$\displaystyle \frac{(1-(-1)^n)}{2}\alpha^{n-1}$
&
$\displaystyle \frac{\psi}{1-\alpha^{2}\psi^{2}}$
&
$\displaystyle 1-\frac{2mr^2}{r^{D-1}
+\sqrt{r^{2(D-1)}+4\alpha^2 m^2}}$ \\

$(e)$
&
$\displaystyle n\,\alpha^{n-1}$
&
$\displaystyle \frac{\psi}{(1-\alpha\psi)^{2}}$
&
$\displaystyle 1-\frac{2mr^2}
{r^{D-1}+2\alpha m+\sqrt{r^{2(D-1)}+4m\alpha r^{D-1}}}$ \\

$(f)$
&
$\displaystyle \frac{\Gamma(2n-1)\alpha^{n-1}}
{4^{n-1}\Gamma(n)^2}$
&
$\displaystyle \frac{\psi}{\sqrt{1-\alpha\psi}}$
&
$\displaystyle 1-\frac{2mr^2}
{m\alpha+\sqrt{4r^{2(D-1)}+m^2\alpha^2}}$ \\

\hline
\end{tabular}
\end{table}


\begin{table}[t]
\centering
\caption{Extension of Table~\ref{tab:solutions}. We display the mass parameter $m$ expressed in terms of the event horizon radius $r_0$, which is determined by the condition $f(r_0)=0$ and depends on the specific model. We also show the corresponding constraints on the coupling parameter $\alpha$ for the regular black hole configurations considered in this work.}
\label{tab:ext-tab-sol}
\setlength{\tabcolsep}{4pt}
\renewcommand{\arraystretch}{1.3}
\begin{tabular}{c c c}
\hline
Label & \makecell[c]{$m$ parameter in\\horizon units}
& \makecell[c]{Constraint\\in $\alpha$} \\
\hline

$(a)$
&
$\displaystyle \frac{r_0^{D-1}}{r_0^2-\alpha}$
&
$\displaystyle 0 \le \alpha \le \frac{D-3}{D-1}\,r_0^2$ \\

$(b)$
&
$\displaystyle \frac{r_0^{D-1}}{\sqrt{r_0^4-\alpha^2}}$
&
$\displaystyle 0 \le \alpha \le \sqrt{\frac{D-3}{D-1}}\,r_0^2$ \\

$(c)$
&
$\displaystyle \frac{r_0^{D-1}}{\alpha}
\ln\!\left(\frac{r_0^2}{r_0^2-\alpha}\right)$
&
--- \\

$(d)$
&
$\displaystyle \frac{r_0^{D+1}}{r_0^4-\alpha^2}$
&
$\displaystyle 0 \le \alpha \le \sqrt{\frac{D-3}{D+1}}\,r_0^2$ \\

$(e)$
&
$\displaystyle \frac{r_0^{D+1}}{(r_0^2-\alpha)^2}$
&
$\displaystyle 0 \le \alpha \le \frac{D-3}{D+1}\,r_0^2$ \\

$(f)$
&
$\displaystyle \frac{r_0^{D-2}}{\sqrt{r_0^2-\alpha}}$
&
$\displaystyle 0 \le \alpha \le \frac{D-3}{D-2}\,r_0^2$ \\

\hline
\end{tabular}
\end{table}

In general, a fundamental higher-curvature theory would be expected to impose specific relations among the coefficients $\alpha_n$. However, in the absence of a complete underlying theory, such relations are not uniquely determined. Instead, one typically considers particular choices of the coefficients $\alpha_n$ as illustrative examples that lead to well-defined and tractable models. In this spirit, several explicit constructions were proposed in \cite{Bueno:2024dgm}, where different ad hoc relations among the $\alpha_n$ define infinite series with desirable properties, such as convergence and the existence of regular black hole solutions. These models, labeled $(a)$–$(e)$, should therefore be understood as representative examples rather than derivations from first principles.

In the same vein, we introduce an additional model $(f)$ corresponding to a different choice of the coefficients $\alpha_n$, defined by
\begin{equation}
\alpha_n=\frac{\Gamma(2n-1)\alpha^{n-1}}{4^{n-1}\Gamma(n)^2}.
\end{equation}
As in the previous cases, this construction is not derived from an underlying fundamental principle, but rather constitutes another ad hoc assignment of the couplings. Its interest lies in the fact that the series in Eq.~\ref{eq:hpsi} can be resummed into a simple closed form, ensuring both convergence and analytical control. Furthermore, this choice guarantees that $\psi>0$ in the relevant domain, so that the characteristic function is monotonic and therefore invertible.

Substituting this expression into the series in Eq.~\ref{eq:hpsi}, one obtains the characteristic function
\begin{equation}
h(\psi)=\frac{\psi}{\sqrt{1-\alpha \psi}}.
\end{equation}
Then, using the solution derived from the second equation of motion, Eq.~\ref{eq:f(r)}, the metric function takes the form
\begin{equation}
f(r)= 1-\frac{2 m r^2}{m\alpha + \sqrt{4 r^{2(D-1)}+m^2\alpha^2}}.
\end{equation}

The resulting regular black hole solution is of differentiability class $C^{2D-1}$ at $r=0$.

The spacetime exhibits two horizons, namely an outer (event) horizon $r_0$ and an inner horizon $r_{-}$, together with a regular de Sitter core at the origin. As the ratio $\alpha/r_0^2$ increases, the inner horizon approaches the event horizon, and both coincide at a critical value corresponding to an extremal configuration. This behavior is qualitatively similar to that of charged black hole solutions, where an additional parameter controls the approach to extremality.

In the present work, we investigate the dynamics of test fields propagating in the background of regular black-hole spacetimes, rather than gravitational perturbations of the underlying higher-curvature theory. The latter constitute a considerably more challenging problem, since they require perturbing all higher-curvature terms entering the infinite series and consistently resumming their contributions to the linearized field equations. In contrast, the test fields considered here couple only to the fixed background geometry. Nevertheless, their quasinormal spectra provide important characteristics of the black-hole spacetime, since they probe the shape of the effective potential, the photon-sphere region, and the way in which regularity modifies wave propagation. In particular, we focus on scalar and electromagnetic perturbations, governed respectively by the Klein–Gordon and Maxwell equations in $D$-dimensional spacetime,
\begin{subequations}\label{eq:fields}
\begin{align}\label{KleinGordon}
\frac{1}{\sqrt{-g}} \partial_{\mu} \Big( \sqrt{-g} \, g^{\mu\nu} \partial_{\nu} \Phi \Big) &= 0, \\\label{Maxwell}
\frac{1}{\sqrt{-g}} \partial_\mu \Big( \sqrt{-g} \, F_{\rho\sigma} g^{\rho\nu} g^{\sigma\mu} \Big) &= 0,
\end{align}
\end{subequations}
where \(\Phi\) is the scalar field and \(F_{\mu\nu} = \partial_\mu A_\nu - \partial_\nu A_\mu\) is the field tensor derived from the vector potential \(A_\mu\). A separation of variables is carried out, leading to a Schrödinger-type wave equation
\begin{equation}\label{eq:wavelike}
\left( \frac{d^2}{dr_*^2} + \omega^2 - V(r_*) \right) \Psi(r_*) = 0,
\end{equation}
where \(r_*\) is the tortoise coordinate defined by
\begin{equation}
\frac{dr_*}{dr} \equiv \frac{1}{f(r)},
\qquad
r_* \to
\begin{cases}
-\infty, & r \to r_0, \\[4pt]
+\infty, & r \to \infty.
\end{cases}
\end{equation}
and \(V(r_*)\) is the corresponding effective potential.

The effective potential for a massless scalar field, \(V_s(r)\), arises after separation of variables in Eq.~(\ref{KleinGordon}). For the electromagnetic field (\ref{Maxwell}), Feynman's gauge is employed (as in \cite{Crispino:2000jx}) to obtain the effective potentials \(V_1(r)\) and \(V_2(r)\) and following the notation of \cite{Lopez-Ortega:2006vjp}:

\begin{subequations}
\begin{equation}
    V_{s}(r) = f(r) \left( \frac{\ell(\ell+D-3)}{r^2} + \right.\frac{(D-2)(D-4)}{4r^2}f(r)
    \left. + \frac{D-2}{2r}\frac{df}{dr}\right),
    \label{eq:escalarpot}
\end{equation}
\begin{equation}
    V_{1}(r) = f(r) \left( \frac{\ell(\ell+D-3)}{r^2} + \right.\frac{(D-2)(D-4)}{4r^2}f(r)
    \left. - \frac{D-4}{2r}\frac{df}{dr}\right),
\end{equation}
\begin{equation}
    V_{2}(r) = f(r) \left( \frac{(\ell+1)(\ell+D-4)}{r^2} + \right.\frac{(D-4)(D-6)}{4r^2}f(r)
    \left. + \frac{D-4}{2r}\frac{df}{dr}\right),
\end{equation}
\end{subequations}

To study the dynamical response of the test fields on the background, we impose QNM boundary conditions on the wavelike equation~(\ref{eq:wavelike}),
\begin{equation}\label{eq:qnmbc}
    \Psi(r_*\to\pm\infty)\propto e^{\pm i\omega r_*}.
\end{equation}
These boundary conditions are imposed to ensure that perturbations behave as purely ingoing waves at the event horizon and as purely outgoing waves at spatial infinity, in accordance with the causal structure of black hole spacetimes. The corresponding eigenvalue problem admits a discrete spectrum of complex frequencies, whose real parts encode the oscillatory behavior of the perturbations, while the imaginary parts characterize their damping rates. We investigate how the QNM spectrum depends on the coupling parameter $\alpha$ across the six families of regular black holes summarized in Tables~\ref{tab:solutions} and \ref{tab:ext-tab-sol}. Our primary goal is to clarify how higher-curvature corrections, responsible for the emergence of regular geometries, modify the spectrum in comparison to the solutions in General Relativity. Furthermore, we perform a systematic comparison of the quasinormal spectra between different models, allowing us to assess the extent to which distinct regularization mechanisms deform the underlying spacetime and leave observable imprints in the ringdown signal.




\section{WKB method}\label{sec:WKB}

To determine the quasinormal modes (QNMs), we employ the WKB method, which is well suited for smooth potentials with a single, well-defined peak, where $\omega^2$ plays the role of an effective energy, and an analogy with quantum-mechanical scattering near the barrier peak applies. Following the original formulation in \cite{Schutz:1985km}, we use the WKB expansion extended to higher orders: from second to third order obtained in \cite{Iyer:1986np}, from fourth to sixth order as given in \cite{Konoplya:2003ii}, and from seventh to thirteenth order as extended in \cite{Matyjasek:2017psv}. Although the expansion has been further derived to seventeenth order~\cite{Matyjasek:2019eeu} and, in principle, can be carried to even higher orders using the Bender–Wu method~\cite{Hatsuda:2019eoj}, in this work we restrict ourselves to the WKB expansion up to thirteenth order. The general expression reads
\begin{equation}
    i\frac{\omega^{2}-V_0}{\sqrt{-2V_0''}} - \sum_{k=2}^{N} \Lambda_{k} \;=\; n + \tfrac{1}{2},
    \label{eq:correctedomega}
\end{equation}
where in the numerical calculations we set $N=13$. Here $V_0$ is the potential evaluated at its maximum and $V_0''$ its second derivative with respect to the tortoise coordinate at $r_{\rm max}$. The higher-order corrections $\Lambda_{k}$ depend on the higher derivatives of the potential at its maximum.

\begin{table}[h]
\centering
\caption{Quasinormal frequencies of test scalar-field perturbations in $D=5$ for configuration $(f)$, with $\alpha = 0.4 r_{0}^{2}$ and overtone number $n=0$, for the monopole ($\ell=0$) and dipole ($\ell=1$) modes.}
\begin{tabular}{c c c c}
\hline
Order & \makecell[c]{Padé\\approximant} & $V_{s}(\ell=0)$ & $V_{s}(\ell=1)$ \\
\hline
7  & $\tilde{m}=4$ & $0.471163 - 0.298991 i$ & $0.909431 - 0.289025 i$ \\
8  & $\tilde{m}=4$ & $0.471307 - 0.297507 i$ & $0.909367 - 0.288982 i$ \\
9  & $\tilde{m}=5$ & $0.471511 - 0.298055 i$ & $0.909358 - 0.288996 i$ \\
10 & $\tilde{m}=5$ & $0.471685 - 0.297529 i$ & $0.909366 - 0.288981 i$ \\
11 & $\tilde{m}=6$ & $0.471648 - 0.298404 i$ & $0.909401 - 0.288962 i$ \\
12 & $\tilde{m}=6$ & $0.471869 - 0.298017 i$ & $0.909394 - 0.288949 i$ \\
13 & $\tilde{m}=7$ & $0.471959 - 0.298046 i$ & $0.909408 - 0.288954 i$ \\
14 & $\tilde{m}=7$ & $0.471869 - 0.298014 i$ & $0.909405 - 0.288954 i$ \\
15 & $\tilde{m}=8$ & $0.472061 - 0.298456 i$ & $0.909408 - 0.288954 i$ \\
16 & $\tilde{m}=8$ & $0.471957 - 0.298362 i$ & $0.909403 - 0.288954 i$ \\
\hline
\end{tabular}
\label{tab:wkb_pade}
\end{table}

\begin{figure}[h]
    \begin{subfigure}[b]{0.5\textwidth}
        \includegraphics[width=\linewidth]{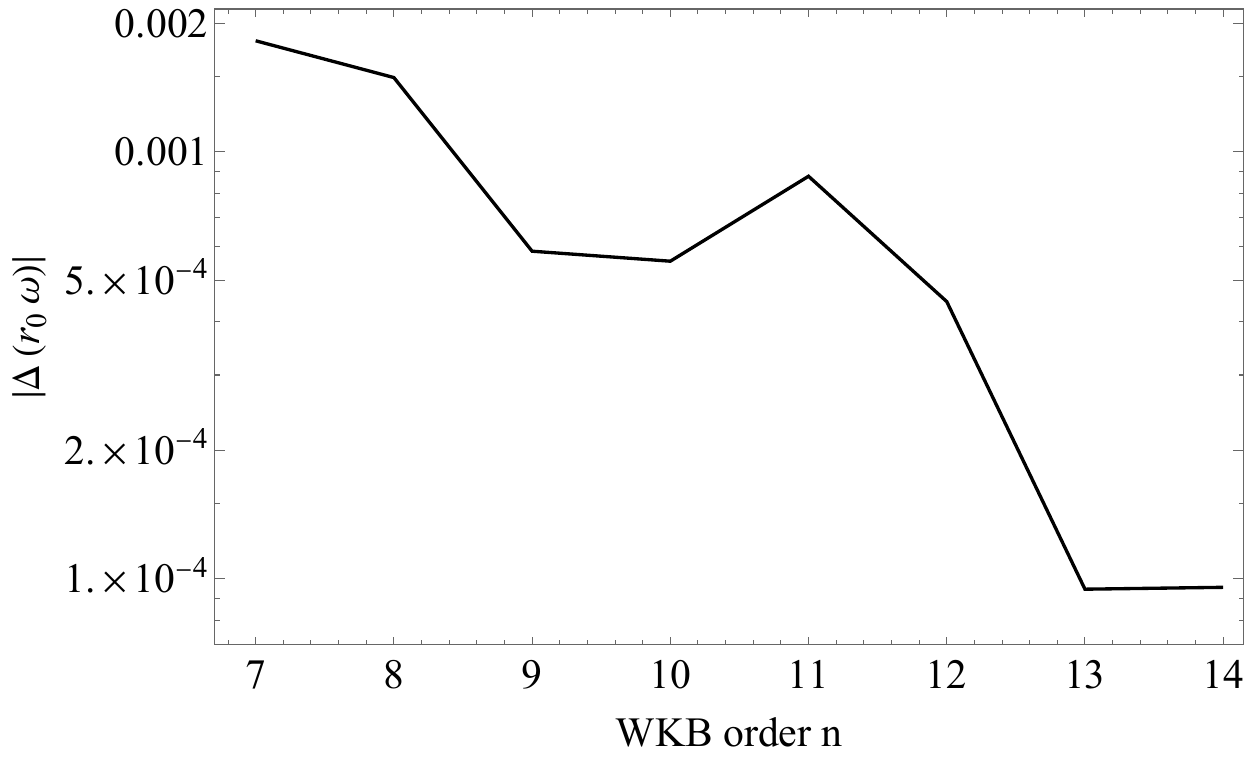}
    \end{subfigure}
    \hfill
    \begin{subfigure}[b]{0.5\textwidth}
        \includegraphics[width=\linewidth]{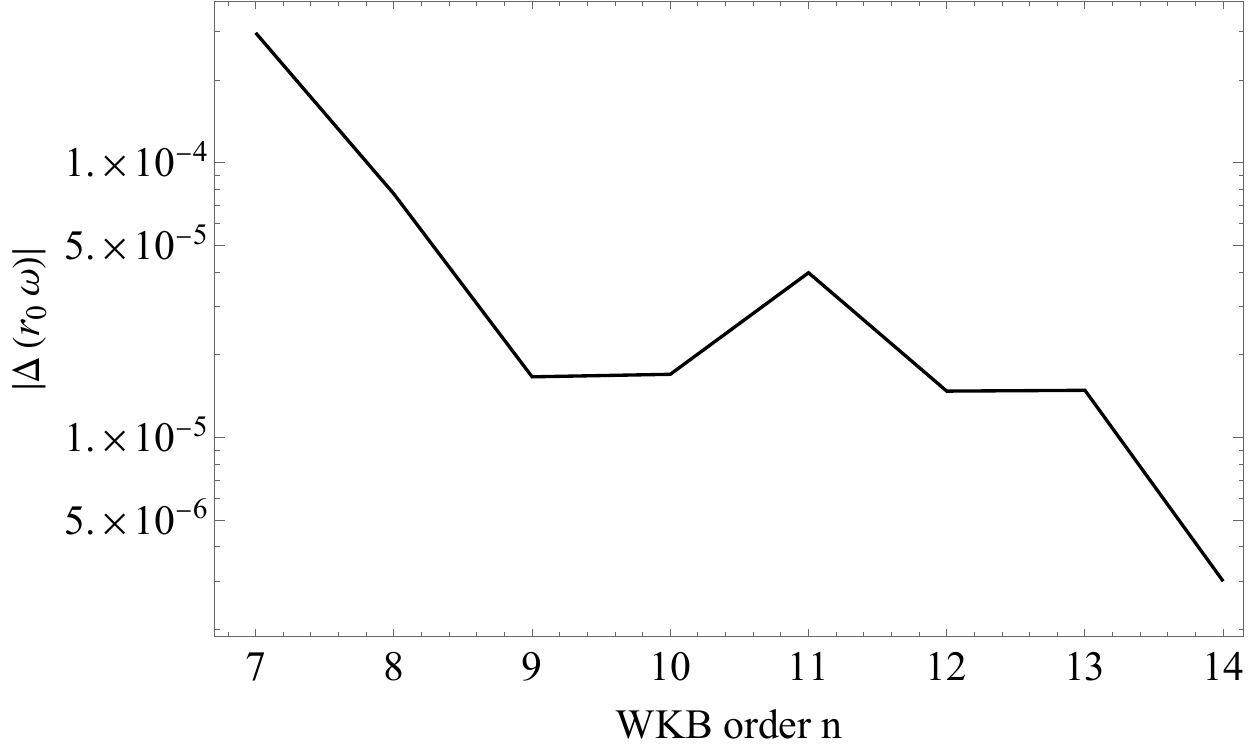}
    \end{subfigure}
    \caption{Absolute difference between consecutive WKB orders, $|\omega_n - \omega_{n-1}|$, for the quasinormal frequencies listed in Table~\ref{tab:wkb_pade}, computed for model $(f)$ with the same parameters as in the table. The left panel corresponds to the monopole mode ($\ell=0$), while the right panel shows the dipole mode ($\ell=1$). The first point corresponds to the difference with the 6th-order $(\tilde{m}=3)$ result.}
    \label{fig:diff}
\end{figure}

To improve convergence, we employ Padé approximants to the WKB expansion \cite{Matyjasek:2017psv,Konoplya:2019hlu}. This approach is naturally motivated by the asymptotic character of the WKB method, which makes the inclusion of Padé approximants particularly effective in enhancing both the accuracy and stability of the results. In Table~\ref{tab:wkb_pade}, we present quasinormal frequencies computed for fixed values of the parameters, with the aim of identifying the order at which the method, supplemented with Padé approximants, reaches a stable regime in comparison with the standard WKB approach without such improvements.

This behavior is clearly illustrated in Fig.~\ref{fig:diff}, where we analyze the absolute differences between consecutive WKB orders. For the monopole case ($\ell=0$), which is the most delicate one for WKB methods, the spread of the last several Padé-improved orders provides an estimate of the residual numerical uncertainty. For the representative model $(f)$ with $D=5$ and $\alpha=0.4r_0^2$, the real part varies by about $7\times10^{-4}$ and the imaginary part by about $9\times10^{-4}$ between the 11th and 14th orders, while the corresponding dipole variations are below $5\times10^{-5}$. Thus the monopole data should be read with lower precision than the higher-multipole data, but the monotonic trend with $\alpha$ is stable under changes of WKB and Padé order. Similarly, for the dipole case ($\ell=1$), the sequence also exhibits convergence toward stable values from approximately the 11th order onward. Although the 13th-order WKB result with Padé approximants yields an accurate estimate, the improvement obtained by going to higher orders becomes marginal.

Furthermore, the inclusion of higher-order terms significantly increases the computational cost, particularly when considering parameter regimes closer to extremality in $\alpha/r_{0}^{2}$ or higher spacetime dimensions. Therefore, the 13th-order WKB method with Padé approximants represents an optimal compromise between accuracy and computational efficiency. While the present analysis has been performed for scalar perturbations of test fields, a similar pattern of asymptotic convergence and stabilization is expected to hold for electromagnetic perturbations.

By comparing the intermediate Padé values at different orders, we grouped those with the lowest variance. Among the possible combinations, the 13th-order expansion with $\tilde{m}=7$ provided the most stable results, while the 9th ($\tilde{m}=5$) and 11th ($\tilde{m}=6$) orders also yielded viable approximations. Therefore, the WKB approach requires the computation of a large number of higher-order derivatives of the effective potential, rendering their determination increasingly expensive and susceptible to numerical loss of precision. This computational burden grows rapidly with the spacetime dimension $D$ and constitutes a practical limitation on the achievable WKB order, particularly when exploring extended regions of the parameter space or analyzing families of regular black hole metrics.

It should be emphasized that the WKB accuracy deteriorates in the regime of large $D$ most notably for the monopole case $\ell = 0$, which is well known to yield the poorest WKB approximations \cite{Kodama:2009bf}. For this reason, the scalar monopole frequencies in the tables are mainly used to display qualitative trends, whereas quantitative conclusions are based preferentially on the more stable $\ell\geq1$ modes and on the large-$\ell$ analytic expansion. In contrast, for higher multipoles ($\ell > 1$), the accuracy of the WKB method significantly improves, especially when supplemented with Padé approximants \cite{Konoplya:2025uiq}. Thus, the higher-order WKB formula with Padé resummation provides reliable results and has been widely used in recent years for quasinormal mode calculations \cite{Hatsuda:2019eoj,Eniceicu:2019npi,Lin:2022ynv,Hatsuda:2023geo,Miyachi:2025ptm}, particularly in cases where direct numerical methods are cumbersome \cite{Panotopoulos:2019gtn,Skvortsova:2024atk,Churilova:2019qph,Konoplya:2019xmn,Konoplya:2021ube,Becar:2022wcj,Bolokhov:2023bwm,Davey:2023fin,Malik:2023bxc,Zinhailo:2024kbq,Dubinsky:2024hmn,Gingrich:2024tuf,Bolokhov:2024ixe,Skvortsova:2023zca,Livine:2024bvo,Dubinsky:2024mwd,Konoplya:2024lch,Cavalcante:2024kmy,Malik:2024nhy,Zhu:2024wic,Dubinsky:2024nzo,Dubinsky:2024vbn,Davlataliev:2024mjl,Stuchlik:2025ezz,Skvortsova:2023zmj,Lutfuoglu:2025hwh}. For higher-dimensional regular black holes, the WKB approximation has been compared with accurate results obtained via the continued fraction method, demonstrating good agreement \cite{Konoplya:2024hfg}.

In addition, the WKB formula helps to understand a particularly important link between classical and perturbative dynamics, which is the correspondence between unstable null geodesics and quasinormal modes in the eikonal regime. In this limit, the effective potential governing perturbations is sharply peaked near the photon sphere, so that the dominant quasinormal oscillations are governed by the properties of the unstable circular null orbit. As shown in ~\cite{Cardoso:2008bp,Dolan:2010wr}, the eikonal quasinormal frequencies ($\ell \gg 1$) can be expressed in terms of the orbital frequency $\Omega_{c}$ and the Lyapunov exponent $\lambda$ of the unstable photon orbit as
\begin{equation}
\omega_{n} \approx \ell \Omega_{c} - i\left(n+\frac{1}{2}\right)\lambda, \qquad  n=0,1,2,\ldots.
\end{equation}
Here $\Omega_{c}$ sets the real oscillation frequency, while $\lambda$ determines the instability timescale and thus the damping rate. This correspondence provides a geometric interpretation of the eikonal quasinormal spectrum and establishes a direct connection between wave dynamics and null geodesics of the underlying spacetime. However, as was shown in \cite{Khanna:2016yow,Konoplya:2017wot,Konoplya:2022gjp,Bolokhov:2023dxq,Konoplya:2025afm,Konoplya:2020bxa} the correspondence can break down in modified theories or for perturbation sectors whose eikonal potentials are not governed by the same null geodesics, so a case-by-case check is necessary. For the test scalar and electromagnetic fields considered here, the leading eikonal part of each effective potential is $V(r)\simeq f(r)\ell^2/r^2$. Its maximum therefore satisfies the photon-sphere condition
\begin{equation}\label{eq:photonsphere}
    2f(r_c)-r_c f'(r_c)=0,
\end{equation}
where $r_c$ denotes the photon-sphere radius and the corresponding orbital frequency and Lyapunov exponent are
\begin{equation}\label{eq:photonobservables}
    \Omega_c=\frac{\sqrt{f(r_c)}}{r_c},
    \qquad
    \lambda=\sqrt{\frac{f(r_c)\left[2f(r_c)-r_c^2 f''(r_c)\right]}{2r_c^2}}.
\end{equation}
For example, for model $(f)$ with $D=5$, Eq.~(\ref{eq:photonsphere}) gives, to quadratic order in $\alpha/r_0^2$,
\begin{equation}\label{eq:photonradius}
    r_c=\sqrt{2}\,r_0\left[1+\frac{3\alpha}{16r_0^2}+\mathcal{O}(\alpha^2)\right].
\end{equation}
Substituting Eq.~(\ref{eq:photonradius}) into Eq.~(\ref{eq:photonobservables}), one obtains
\begin{equation}
\Omega_c=\frac{1}{2r_0}\left[1-\frac{3\alpha}{16r_0^2}+\mathcal{O}(\alpha^2)\right],\qquad \lambda =\frac{1}{\sqrt{2}r_0}\left[1-\frac{7\alpha}{16r_0^2}+\mathcal{O}(\alpha^2)\right].
\end{equation}
These coefficients reproduce the leading real and imaginary parts of the large-$\ell$ expansion displayed below, confirming the eikonal correspondence for the perturbations analyzed in this work.

The quasinormal modes studied here could also be used to obtain grey-body factors via the correspondence established in~\cite{Konoplya:2024lir},
\begin{equation}
\Gamma_{\ell}(\Omega) \approx \left[ 1 + \exp\left( \frac{2\pi\bigl(\Omega^{2}-\mathrm{Re}(\omega_{0})^{2}\bigr)}{4\,\mathrm{Re}(\omega_{0})\,\mathrm{Im}(\omega_{0})} \right) \right]^{-1}.
\end{equation}
Here $\Omega$ is the real frequency of the radiation propagating under the scattering boundary conditions and $\omega_0$ is the least-damped quasinormal frequency ($n=0$). This correspondence was tested and applied in a number of recent publications for a variety of black-hole backgrounds \cite{Lutfuoglu:2025ldc,Malik:2024cgb,Skvortsova:2024msa,Lutfuoglu:2025blw,Bolokhov:2024otn,Han:2025cal,Dubinsky:2025nxv,Bolokhov:2025lnt,Malik:2025dxn}.

\section{Quasinormal modes of regular black holes}\label{sec:qnms}

\begin{table}[t]
\centering

\caption{Dominant ($n=0$) QNMs of the scalar and electromagnetic fields for $D=5,\dots,8$, with $r_0=1$, computed using the 13th-order WKB method with Pad\'e approximant ($\tilde{m}=7$) for model (c).}
\label{table:c}
\begin{tabular}{c c c c c}
\hline
$\alpha$ & $V_{s}(\ell=0)$ & $V_{s}(\ell=1)$ & $V_{1}(\ell=1)$ & $V_{2}(\ell=1)$ \\
\hline

\multicolumn{5}{c}{$D=5$} \\
\hline
0.1 & 0.524612 - 0.363382$i$ & 0.996278 - 0.344929$i$ & 0.746266 - 0.300502$i$ & 0.938018 - 0.334019$i$ \\
0.2 & 0.512551 - 0.342817$i$ & 0.973320 - 0.327006$i$ & 0.734760 - 0.284687$i$ & 0.920196 - 0.316326$i$ \\
0.3 & 0.496812 - 0.321422$i$ & 0.946526 - 0.308891$i$ & 0.719657 - 0.267636$i$ & 0.898469 - 0.297895$i$ \\
0.4 & 0.476671 - 0.300767$i$ & 0.915315 - 0.291094$i$ & 0.700063 - 0.249979$i$ & 0.871931 - 0.279252$i$ \\
0.5 & 0.451927 - 0.284005$i$ & 0.879304 - 0.274122$i$ & 0.675073 - 0.232707$i$ & 0.839799 - 0.261201$i$ \\
0.6 & 0.427739 - 0.269589$i$ & 0.837991 - 0.257882$i$ & 0.644384 - 0.216725$i$ & 0.801502 - 0.244319$i$ \\
0.7 & 0.402230 - 0.253655$i$ & 0.789672 - 0.241674$i$ & 0.607525 - 0.202222$i$ & 0.755700 - 0.228349$i$ \\
0.71 & 0.399494 - 0.251931$i$ & 0.784327 - 0.240015$i$ & 0.603418 - 0.200821$i$ & 0.750592 - 0.226771$i$ \\

\hline
\multicolumn{5}{c}{$D=6$} \\
\hline
0.1 & 0.878142 - 0.51249$i$ & 1.42661 - 0.491152$i$ & 1.03061 - 0.431934$i$ & 1.38437 - 0.480591$i$ \\
0.2 & 0.863919 - 0.491044$i$ & 1.40355 - 0.472404$i$ & 1.01740 - 0.412542$i$ & 1.36579 - 0.461908$i$ \\
0.3 & 0.845995 - 0.468789$i$ & 1.37671 - 0.453257$i$ & 0.999737 - 0.400145$i$ & 1.34344 - 0.442310$i$ \\
0.4 & 0.823498 - 0.446759$i$ & 1.34541 - 0.434076$i$ & 0.974079 - 0.382315$i$ & 1.31641 - 0.422102$i$ \\
0.5 & 0.795666 - 0.425857$i$ & 1.30902 - 0.415304$i$ & 0.946766 - 0.360973$i$ & 1.28369 - 0.401805$i$ \\
0.6 & 0.765041 - 0.409079$i$ & 1.26684 - 0.396959$i$ & 0.916230 - 0.340799$i$ & 1.24423 - 0.381938$i$ \\
0.7 & 0.732589 - 0.391954$i$ & 1.21705 - 0.378303$i$ & 0.880029 - 0.322070$i$ & 1.19633 - 0.362421$i$ \\
0.8 & 0.694442 - 0.371509$i$ & 1.15487 - 0.357837$i$ & 0.834932 - 0.303660$i$ & 1.13554 - 0.342199$i$ \\

\hline
\multicolumn{5}{c}{$D=7$} \\
\hline
0.1 & 1.25769 - 0.644654$i$ & 1.86099 - 0.622420$i$ & 1.39237 - 0.597216$i$ & 1.82926 - 0.612691$i$ \\
0.2 & 1.24164 - 0.622715$i$ & 1.83742 - 0.603142$i$ & 1.36728 - 0.578443$i$ & 1.80975 - 0.593376$i$ \\
0.3 & 1.22171 - 0.600070$i$ & 1.81004 - 0.583392$i$ & 1.33575 - 0.540187$i$ & 1.78649 - 0.573093$i$ \\
0.4 & 1.19733 - 0.577308$i$ & 1.77814 - 0.563502$i$ & 1.30994 - 0.515879$i$ & 1.75855 - 0.552090$i$ \\
0.5 & 1.16722 - 0.555222$i$ & 1.74098 - 0.543865$i$ & 1.28506 - 0.495907$i$ & 1.72489 - 0.530769$i$ \\
0.6 & 1.13290 - 0.536349$i$ & 1.69775 - 0.524550$i$ & 1.25509 - 0.477045$i$ & 1.68429 - 0.509541$i$ \\
0.7 & 1.09535 - 0.518249$i$ & 1.64658 - 0.504812$i$ & 1.21712 - 0.457539$i$ & 1.63481 - 0.488307$i$ \\
0.8 & 1.05119 - 0.497151$i$ & 1.58243 - 0.482946$i$ & 1.16826 - 0.436345$i$ & 1.57173 - 0.465900$i$ \\
0.85 & 1.02395 - 0.484217$i$ & 1.54174 - 0.470186$i$ & 1.13785 - 0.424570$i$ & 1.53141 - 0.453392$i$ \\

\hline
\multicolumn{5}{c}{$D=8$} \\
\hline
0.1 & 1.65466 - 0.76422$i$ & 2.29980 - 0.741933$i$ & 1.77236 - 0.710614$i$ & 2.27514 - 0.733194$i$ \\
0.2 & 1.63704 - 0.741951$i$ & 2.27561 - 0.722242$i$ & 1.74698 - 0.698635$i$ & 2.25477 - 0.713276$i$ \\
0.3 & 1.61550 - 0.718929$i$ & 2.24756 - 0.702111$i$ & 1.72036 - 0.683636$i$ & 2.23053 - 0.692602$i$ \\
0.4 & 1.58968 - 0.695395$i$ & 2.21489 - 0.681800$i$ & 1.69104 - 0.664782$i$ & 2.20162 - 0.671174$i$ \\
0.5 & 1.55763 - 0.673217$i$ & 2.17690 - 0.661720$i$ & 1.65848 - 0.646065$i$ & 2.16695 - 0.649338$i$ \\
0.6 & 1.52068 - 0.653135$i$ & 2.13263 - 0.641921$i$ & 1.62236 - 0.627411$i$ & 2.12521 - 0.627459$i$ \\
0.7 & 1.47938 - 0.634315$i$ & 2.08019 - 0.621700$i$ & 1.58048 - 0.608073$i$ & 2.07436 - 0.605395$i$ \\
0.8 & 1.43054 - 0.612916$i$ & 2.01437 - 0.599221$i$ & 1.52869 - 0.586577$i$ & 2.00946 - 0.581881$i$ \\
0.88 & 1.37894 - 0.59060$i$ & 1.94248 - 0.576938$i$ & 1.47343 - 0.565037$i$ & 1.93795 - 0.559718$i$ \\
\hline
\end{tabular}
\end{table}

\begin{table}[t]
\centering
\caption{Dominant ($n=0$) QNMs of the scalar and electromagnetic fields for $D=5,\dots,8$, with $r_0=1$, computed using the 13th-order WKB method with Pad\'e approximant ($\tilde{m}=7$) for model (d).}
\label{table:d}

\begin{tabular}{c c c c c}
\hline
$\alpha$ & $V_{s}(\ell=0)$ & $V_{s}(\ell=1)$ & $V_{1}(\ell=1)$ & $V_{2}(\ell=1)$ \\
\hline

\multicolumn{5}{c}{$D=5$} \\
\hline
0   & 0.533578 - 0.383171$i$ & 1.01602 - 0.362328$i$ & 0.755120 - 0.315841$i$ & 0.952727 - 0.350740$i$ \\
0.1 & 0.531257 - 0.378356$i$ & 1.01073 - 0.358338$i$ & 0.752230 - 0.311877$i$ & 0.948617 - 0.346605$i$ \\
0.2 & 0.522473 - 0.363749$i$ & 0.994335 - 0.346933$i$ & 0.742876 - 0.300064$i$ & 0.935586 - 0.334584$i$ \\
0.3 & 0.504863 - 0.341767$i$ & 0.965592 - 0.329927$i$ & 0.725107 - 0.281553$i$ & 0.911596 - 0.316129$i$ \\
0.4 & 0.475015 - 0.318784$i$ & 0.924016 - 0.310255$i$ & 0.696832 - 0.259603$i$ & 0.874726 - 0.294446$i$ \\
0.5 & 0.442757 - 0.299810$i$ & 0.870306 - 0.289359$i$ & 0.657313 - 0.238119$i$ & 0.824814 - 0.272294$i$ \\

\hline
\multicolumn{5}{c}{$D=6$} \\
\hline
0    & 0.889469 - 0.532893$i$ & 1.44650 - 0.509272$i$ & 1.05229 - 0.469168$i$ & 1.39993 - 0.498228$i$ \\
0.1  & 0.886133 - 0.527960$i$ & 1.44114 - 0.505240$i$ & 1.03997 - 0.449490$i$ & 1.39557 - 0.494014$i$ \\
0.2  & 0.875662 - 0.513451$i$ & 1.42456 - 0.493642$i$ & 1.02638 - 0.432106$i$ & 1.38186 - 0.481713$i$ \\
0.3  & 0.855374 - 0.491047$i$ & 1.39553 - 0.475999$i$ & 1.00796 - 0.419975$i$ & 1.35494 - 0.459818$i$ \\
0.4  & 0.822283 - 0.465298$i$ & 1.35310 - 0.454956$i$ & 0.977428 - 0.402928$i$ & 1.31902 - 0.438982$i$ \\
0.5  & 0.781231 - 0.443416$i$ & 1.29763 - 0.432059$i$ & 0.897435 - 0.304044$i$ & 1.26676 - 0.413475$i$ \\
0.51 & 0.777012 - 0.441192$i$ & 1.29134 - 0.429653$i$ & 0.893961 - 0.302554$i$ & 1.26072 - 0.410876$i$ \\

\hline
\multicolumn{5}{c}{$D=7$} \\
\hline
0   & 1.27066 - 0.665679$i$ & 1.88139 - 0.641058$i$ & 1.41701 - 0.609161$i$ & 1.84576 - 0.630977$i$ \\
0.1 & 1.26695 - 0.660627$i$ & 1.87586 - 0.637015$i$ & 1.41620 - 0.603769$i$ & 1.84113 - 0.626742$i$ \\
0.2 & 1.25492 - 0.646155$i$ & 1.85880 - 0.625361$i$ & 1.39650 - 0.583845$i$ & 1.82665 - 0.614339$i$ \\
0.3 & 1.23254 - 0.623967$i$ & 1.82910 - 0.607521$i$ & 1.35065 - 0.572328$i$ & 1.80078 - 0.594880$i$ \\
0.4 & 1.19685 - 0.597333$i$ & 1.78554 - 0.586018$i$ & 1.30819 - 0.551269$i$ & 1.76130 - 0.570655$i$ \\
0.5 & 1.15034 - 0.573047$i$ & 1.72830 - 0.562264$i$ & 1.26068 - 0.525172$i$ & 1.70716 - 0.543833$i$ \\
0.6 & 1.09784 - 0.548156$i$ & 1.65581 - 0.535653$i$ & 1.20239 - 0.497023$i$ & 1.63658 - 0.515254$i$ \\
0.7 & 1.03451 - 0.516911$i$ & 1.56233 - 0.504275$i$ & 1.13124 - 0.466301$i$ & 1.54439 - 0.483885$i$ \\

\hline
\multicolumn{5}{c}{$D=8$} \\
\hline
0    & 1.66894 - 0.785539$i$ & 2.32074 - 0.760994$i$ & 1.79597 - 0.721065$i$ & 2.29260 - 0.751917$i$ \\
0.1  & 1.66478 - 0.780600$i$ & 2.31506 - 0.756922$i$ & 1.79580 - 0.716199$i$ & 2.28773 - 0.747673$i$ \\
0.2  & 1.65141 - 0.766041$i$ & 2.29752 - 0.745326$i$ & 1.81079 - 0.745671$i$ & 2.27252 - 0.735169$i$ \\
0.3  & 1.62719 - 0.743678$i$ & 2.26697 - 0.727401$i$ & 1.73355 - 0.685527$i$ & 2.24559 - 0.715715$i$ \\
0.4  & 1.58964 - 0.717338$i$ & 2.22229 - 0.705811$i$ & 1.70266 - 0.682586$i$ & 2.20490 - 0.691184$i$ \\
0.5  & 1.53979 - 0.691801$i$ & 2.16336 - 0.681715$i$ & 1.64957 - 0.669621$i$ & 2.14885 - 0.664060$i$ \\
0.6  & 1.48118 - 0.666100$i$ & 2.08876 - 0.654572$i$ & 1.58178 - 0.647380$i$ & 2.07588 - 0.634488$i$ \\
0.7  & 1.41032 - 0.634713$i$ & 1.99213 - 0.622513$i$ & 1.50079 - 0.617026$i$ & 1.98010 - 0.601404$i$ \\
0.8  & 1.31585 - 0.592163$i$ & 1.85842 - 0.580881$i$ & 1.40080 - 0.575658$i$ & 1.84718 - 0.561367$i$ \\
0.85 & 1.25219 - 0.563230$i$ & 1.76654 - 0.553235$i$ & 1.33637 - 0.547462$i$ & 1.75571 - 0.535892$i$ \\

\hline
\end{tabular}
\end{table}


\begin{table}[t]
\centering
\caption{Dominant ($n=0$) QNMs of the scalar and electromagnetic fields for $D=5,\dots,8$, with $r_0=1$, computed using the 13th-order WKB method with Pad\'e approximant ($\tilde{m}=7$) for model (e).}
\label{table:e}

\begin{tabular}{c c c c c}
\hline
$\alpha$ & $V_{s}(\ell=0)$ & $V_{s}(\ell=1)$ & $V_{1}(\ell=1)$ & $V_{2}(\ell=1)$ \\
\hline
\multicolumn{5}{c}{$D=5$} \\
\hline
0    & 0.533578 - 0.383171$i$ & 1.01602 - 0.362328$i$ & 0.755120 - 0.315841$i$ & 0.952727 - 0.350740$i$ \\
0.1  & 0.490214 - 0.310869$i$ & 0.934101 - 0.298745$i$ & 0.713186 - 0.259430$i$ & 0.888903 - 0.288321$i$ \\
0.2  & 0.428689 - 0.258145$i$ & 0.838103 - 0.247710$i$ & 0.650384 - 0.209802$i$ & 0.805651 - 0.235432$i$ \\
0.3  & 0.372034 - 0.223084$i$ & 0.735184 - 0.209968$i$ & 0.573305 - 0.174404$i$ & 0.709354 - 0.197085$i$ \\
0.33 & 0.355961 - 0.213369$i$ & 0.703764 - 0.200578$i$ & 0.548939 - 0.166393$i$ & 0.679176 - 0.188107$i$ \\
\hline
\multicolumn{5}{c}{$D=6$} \\
\hline
0    & 0.889469 - 0.532893$i$ & 1.44650 - 0.509272$i$ & 1.05229 - 0.469168$i$ & 1.39993 - 0.498228$i$ \\
0.1  & 0.838460 - 0.457316$i$ & 1.36438 - 0.442339$i$ & 0.991299 - 0.382398$i$ & 1.33364 - 0.431884$i$ \\
0.2  & 0.769228 - 0.396263$i$ & 1.26835 - 0.385492$i$ & 0.934489 - 0.325859$i$ & 1.24969 - 0.372503$i$ \\
0.3  & 0.697085 - 0.353386$i$ & 1.16258 - 0.339221$i$ & 0.861015 - 0.279258$i$ & 1.15075 - 0.323455$i$ \\
0.4  & 0.627227 - 0.316301$i$ & 1.04958 - 0.301046$i$ & 0.778160 - 0.244344$i$ & 1.04062 - 0.284939$i$ \\
0.42 & 0.613138 - 0.309117$i$ & 1.02615 - 0.294094$i$ & 0.760806 - 0.238544$i$ & 1.01745 - 0.278260$i$ \\
\hline
\multicolumn{5}{c}{$D=7$} \\
\hline
0   & 1.27066 - 0.665679$i$ & 1.88139 - 0.641058$i$ & 1.41701 - 0.609161$i$ & 1.84576 - 0.630977$i$ \\
0.1 & 1.21337 - 0.587943$i$ & 1.79750 - 0.571929$i$ & 1.32179 - 0.511632$i$ & 1.77628 - 0.562036$i$ \\
0.2 & 1.13840 - 0.523060$i$ & 1.69981 - 0.512050$i$ & 1.26066 - 0.465052$i$ & 1.69001 - 0.499310$i$ \\
0.3 & 1.05658 - 0.474541$i$ & 1.59149 - 0.461386$i$ & 1.17428 - 0.417468$i$ & 1.58876 - 0.444934$i$ \\
0.4 & 0.974568 - 0.433511$i$ & 1.47374 - 0.417719$i$ & 1.08498 - 0.371396$i$ & 1.47415 - 0.399234$i$ \\
0.5 & 0.888747 - 0.394020$i$ & 1.34537 - 0.378492$i$ & 0.990096 - 0.334296$i$ & 1.34651 - 0.360554$i$ \\
\hline
\multicolumn{5}{c}{$D=8$} \\
\hline
0    & 1.66894 - 0.785539$i$ & 2.32074 - 0.760994$i$ & 1.79597 - 0.721065$i$ & 2.29260 - 0.751917$i$ \\
0.1  & 1.60645 - 0.706229$i$ & 2.23469 - 0.690227$i$ & 1.65711 - 0.683239$i$ & 2.21987 - 0.681084$i$ \\
0.2  & 1.52656 - 0.639299$i$ & 2.13490 - 0.628499$i$ & 1.61464 - 0.574085$i$ & 2.13088 - 0.616343$i$ \\
0.3  & 1.43731 - 0.587572$i$ & 2.02408 - 0.575311$i$ & 1.54136 - 0.531410$i$ & 2.02707 - 0.559045$i$ \\
0.4  & 1.34688 - 0.543174$i$ & 1.90269 - 0.528300$i$ & 1.44638 - 0.488627$i$ & 1.90905 - 0.509046$i$ \\
0.5  & 1.25016 - 0.500871$i$ & 1.76860 - 0.485090$i$ & 1.34225 - 0.446792$i$ & 1.77594 - 0.465098$i$ \\
0.55 & 1.19830 - 0.479620$i$ & 1.69556 - 0.464211$i$ & 1.28649 - 0.427182$i$ & 1.70279 - 0.444740$i$ \\
\hline
\end{tabular}

\end{table}


\begin{table}[t]
\centering
\caption{Dominant $(n=0)$ QNMs of the scalar and electromagnetic fields with $D=5,\dots,8$, $r_0=1$, using the 13th-order WKB method with Padé approximant $(\tilde{m}=7)$ for the model $(f)$.}
\label{table:f}
\renewcommand{\arraystretch}{0.9}
\setlength{\tabcolsep}{4pt}
\resizebox{\linewidth}{!}{%
\begin{tabular}{c c c c c}
\hline
$\alpha$ & $V_{s}(\ell=0)$ & $V_{s}(\ell=1)$ & $V_{1}(\ell=1)$ & $V_{2}(\ell=1)$ \\
\hline
\multicolumn{5}{c}{$D=5$} \\
\hline
0    & $0.533578-0.383171 i$ & $1.016020-0.362328 i$ & $0.755120-0.315841 i$ & $0.952727-0.350740 i$ \\
0.1  & $0.524476-0.363165 i$ & $0.996027-0.344761 i$ & $0.746109-0.300294 i$ & $0.937816-0.333840 i$ \\
0.2  & $0.511843-0.341960 i$ & $0.972162-0.326350 i$ & $0.734090-0.283928 i$ & $0.919224-0.315597 i$ \\
0.3  & $0.494723-0.319667 i$ & $0.943553-0.307517 i$ & $0.717714-0.265963 i$ & $0.895852-0.296302 i$ \\
0.4  & $0.471959-0.298046 i$ & $0.909408-0.288954 i$ & $0.695733-0.247258 i$ & $0.866459-0.276687 i$ \\
0.5  & $0.445383-0.281708 i$ & $0.869230-0.271197 i$ & $0.667064-0.229167 i$ & $0.830080-0.257805 i$ \\
0.6  & $0.419077-0.265824 i$ & $0.822058-0.253864 i$ & $0.631491-0.212648 i$ & $0.785887-0.240106 i$ \\
0.66 & $0.402057-0.255093 i$ & $0.789314-0.243245 i$ & $0.606438-0.203410 i$ & $0.754733-0.229812 i$ \\
\hline
\multicolumn{5}{c}{$D=6$} \\
\hline
0    & $0.889469-0.532893 i$ & $1.44650-0.509272 i$ & $1.05229-0.469168 i$ & $1.39993-0.498228 i$ \\
0.1  & $0.877975-0.512276 i$ & $1.42635-0.490980 i$ & $1.03021-0.431227 i$ & $1.38416-0.480407 i$ \\
0.2  & $0.863098-0.490167 i$ & $1.40238-0.471731 i$ & $1.01671-0.412376 i$ & $1.36478-0.461158 i$ \\
0.3  & $0.843788-0.466914 i$ & $1.37372-0.451810 i$ & $0.997824-0.399241 i$ & $1.34077-0.440632 i$ \\
0.4  & $0.818678-0.443793 i$ & $1.33941-0.431726 i$ & $0.970771-0.381805 i$ & $1.31083-0.419260 i$ \\
0.5  & $0.787210-0.423043 i$ & $1.29863-0.411991 i$ & $0.939927-0.361508 i$ & $1.27363-0.397774 i$ \\
0.6  & $0.753645-0.405243 i$ & $1.25024-0.392385 i$ & $0.904554-0.341341 i$ & $1.22773-0.376691 i$ \\
0.7  & $0.716324-0.385406 i$ & $1.19109-0.371621 i$ & $0.861713-0.321366 i$ & $1.17032-0.355497 i$ \\
0.75 & $0.694624-0.373738 i$ & $1.15535-0.360167 i$ & $0.835859-0.311166 i$ & $1.13528-0.344360 i$ \\
\hline
\multicolumn{5}{c}{$D=7$} \\
\hline
0    & $1.27066-0.665679 i$ & $1.88139-0.641058 i$ & $1.41701-0.609161 i$ & $1.84576-0.630978 i$ \\
0.1  & $1.25750-0.644440 i$ & $1.86072-0.622249 i$ & $1.39231-0.596857 i$ & $1.82906-0.612500 i$ \\
0.2  & $1.24074-0.621872 i$ & $1.83623-0.602462 i$ & $1.36602-0.574707 i$ & $1.80872-0.592647 i$ \\
0.3  & $1.21923-0.598167 i$ & $1.80699-0.581919 i$ & $1.33230-0.538941 i$ & $1.78374-0.571303 i$ \\
0.4  & $1.19214-0.574266 i$ & $1.77199-0.561082 i$ & $1.30507-0.515378 i$ & $1.75279-0.549068 i$ \\
0.5  & $1.15803-0.551741 i$ & $1.73028-0.540439 i$ & $1.27611-0.494765 i$ & $1.71447-0.526465 i$ \\
0.6  & $1.11951-0.532273 i$ & $1.68060-0.519800 i$ & $1.23987-0.474496 i$ & $1.66719-0.503860 i$ \\
0.7  & $1.07629-0.511741 i$ & $1.61967-0.497818 i$ & $1.19329-0.453011 i$ & $1.64603-0.504676 i$ \\
0.8  & $1.02185-0.485798 i$ & $1.53900-0.471857 i$ & $1.13247-0.428740 i$ & $1.52715-0.454271 i$ \\
\hline
\multicolumn{5}{c}{$D=8$} \\
\hline
0     & $1.66894-0.785539 i$ & $2.32074-0.760994 i$ & $1.79597-0.721065 i$ & $2.29260-0.751917 i$ \\
0.1   & $1.65445-0.764010 i$ & $2.29953-0.741762 i$ & $1.77235-0.710499 i$ & $2.27488-0.733012 i$ \\
0.2   & $1.63599-0.741095 i$ & $2.27438-0.721572 i$ & $1.74650-0.697539 i$ & $2.25369-0.712527 i$ \\
0.3   & $1.61285-0.717046 i$ & $2.24443-0.700652 i$ & $1.71846-0.680623 i$ & $2.22768-0.690916 i$ \\
0.4   & $1.58412-0.692295 i$ & $2.20860-0.679389 i$ & $1.68642-0.661987 i$ & $2.19574-0.668264 i$ \\
0.5   & $1.54786-0.669448 i$ & $2.16594-0.658281 i$ & $1.65082-0.643917 i$ & $2.15635-0.645060 i$ \\
0.6   & $1.50598-0.648844 i$ & $2.11501-0.637161 i$ & $1.61009-0.625466 i$ & $2.10768-0.621665 i$ \\
0.7   & $1.45819-0.627869 i$ & $2.05250-0.614664 i$ & $1.56094-0.605324 i$ & $2.04646-0.597478 i$ \\
0.8   & $1.39784-0.601654 i$ & $1.96953-0.587927 i$ & $1.49683-0.580402 i$ & $1.96418-0.570252 i$ \\
0.83  & $1.37511-0.591850 i$ & $1.93766-0.578261 i$ & $1.47252-0.570994 i$ & $1.93243-0.560774 i$ \\
\hline
\end{tabular}
}
\end{table}

\begin{figure}[t]
    \begin{subfigure}[b]{0.5\textwidth}
        \includegraphics[width=\linewidth]{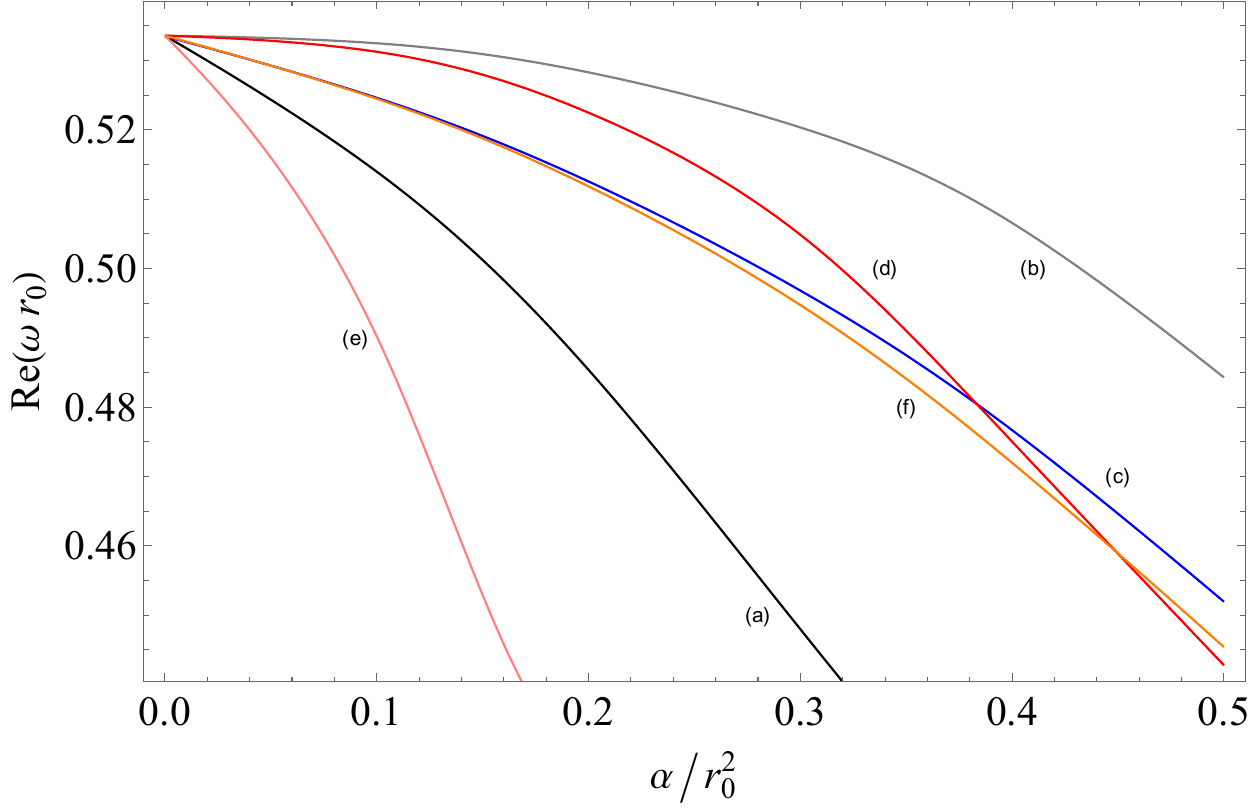}
    \end{subfigure}
    \hfill
    \begin{subfigure}[b]{0.5\textwidth}
        \includegraphics[width=\linewidth]{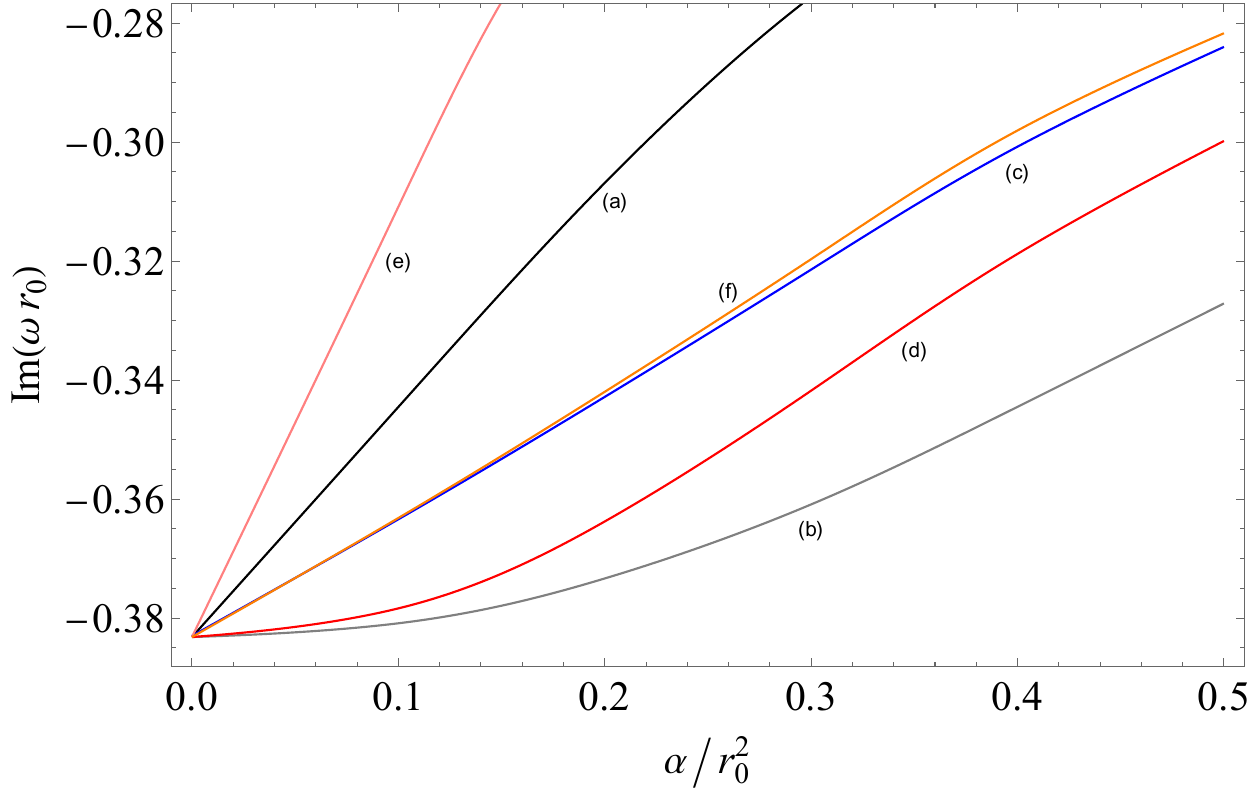}
    \end{subfigure}
    \caption{Dominant ($n=0$) scalar-field QNMs with $\ell=0$ in $D=5$ for various regular black holes: black line $(a)$, gray line $(b)$, blue line $(c)$, red line $(d)$, pink line $(e)$, and orange line $(f)$.}
    \label{fig:graficad5}
\end{figure}

We have computed the QNMs for all the regular black hole models listed in Tables~\ref{tab:solutions} and \ref{tab:ext-tab-sol}. The fundamental modes for models $(c)$, $(d)$, $(e)$, and $(f)$ are presented in Tables~\ref{table:c}, \ref{table:d}, \ref{table:e}, and \ref{table:f}, respectively, while the spectra for models $(a)$ and $(b)$ were previously obtained in \cite{Konoplya:2024hfg}. Our results indicate that the inclusion of higher-curvature corrections, which are responsible for the emergence of regular black hole geometries, produces a qualitatively universal modification of the quasinormal spectrum: both the oscillation frequencies (real parts) and damping rates (absolute values of the imaginary parts) are reduced compared to their general relativity counterparts. The behavior observed in Fig.~\ref{fig:graficad5} for $D=5$ is qualitatively the same for higher dimensions ($D=6,7,8$), where both the real and imaginary parts of the quasinormal frequencies exhibit the same monotonic dependence on the coupling parameter $\alpha$.

In what follows, $r_0$ denotes the radius of the event horizon, defined by the condition $f(r_0)=0$, and serves as the natural length scale of the system. All quantities are expressed in units of $r_0$. In particular, quasinormal frequencies are written in terms of the dimensionless combination $r_0 \omega$, and, when convenient, we set $r_0 = 1$. This normalization is especially well-suited to higher-dimensional spacetimes, where the mass parameter has dimensions $\mathrm{[L]}^{D-3}$ rather than length, so that using $r_0$ provides a consistent and dimension-independent geometric scale.

The deformation of the quasinormal spectrum can also be interpreted from the photon-sphere perspective. In the eikonal regime, the dominant features of the quasinormal modes are governed by the unstable circular null orbit, whose position and instability are modified by the higher-curvature corrections. As the coupling parameter $\alpha$ increases, the regularization effects become stronger and deform the effective potential in the photon-sphere region, shifting the location of its maximum and changing its curvature. This modifies both the orbital frequency and the Lyapunov exponent of the null geodesic, providing a geometric explanation for the observed changes in $\mathrm{Re}(\omega)$ and $|\mathrm{Im}(\omega)|$. Since the same photon-sphere structure also controls the critical impact parameter associated with the black-hole shadow, these corrections may in principle affect the apparent shadow size, although the sign and magnitude of this effect depend on the detailed behavior of $r_c/\sqrt{f(r_c)}$ for each model.

Among the six families, the most pronounced effect occurs for model $(e)$, where the higher-curvature couplings $\alpha_n$ take larger values relative to the other cases (see Fig.~\ref{fig:graficad5}). This can be traced to the structure $h(\psi)=\psi/(1-\alpha\psi)^2$, which weights the higher powers of $\psi$ by the factor $n\alpha^{n-1}$. For fixed horizon radius, this produces a stronger deformation of $f(r)$ in the photon-sphere and potential-barrier region than the milder resummations appearing in models $(c)$, $(d)$, and $(f)$. The maximum of the effective potential is therefore lowered more efficiently, explaining the larger decrease in both $\mathrm{Re}(\omega)$ and $|\mathrm{Im}(\omega)|$. This qualitative trend is also consistent with four-dimensional regular black holes \cite{Toshmatov:2015wga,Rincon:2020cos,Konoplya:2022hll,Konoplya:2023aph}, as well as with the $D$-dimensional generalization of the Dymnikova black hole, such as a regular configuration that cannot be derived within a perturbative expansion in higher-curvature terms yet also exhibits the same suppression of quasinormal frequencies \cite{Konoplya:2024kih}. This suggests that such a suppression may be a generic feature of regular black holes, largely insensitive to the specific mechanism by which regularity is achieved.

The new model $(f)$ occupies an intermediate position among the previously studied cases. Its resummed function $h(\psi)=\psi/\sqrt{1-\alpha\psi}$ produces weaker deviations than model $(e)$ while producing deviations comparable to, and in some parameter ranges slightly stronger than, those of models $(c)$ and $(d)$. This supports the interpretation that the reduction of the test-field frequencies is robust under changes of the resummation prescription.

In order to check whether the observed behavior of the quasinormal spectrum persists for higher multipole numbers, we employ the analytic method developed beyond the eikonal approximation in \cite{Konoplya:2023moy}. The starting point is to rewrite the angular eigenvalue contribution in terms of
\begin{equation}
    \kappa \equiv \ell+\frac{1}{2}(D-3),
\end{equation}
so that the potential admits an expansion of the form
\begin{equation}
    V(r)=\kappa^2 V_{(0)}(r)+V_{(1)}(r)+\mathcal{O}(\kappa^{-2}).
\end{equation}
The leading peak position is obtained from $V_{(0)}'(r)=0$, which coincides with the photon-sphere condition, while subleading corrections follow by solving $V'(r)=0$ order by order in $\kappa^{-1}$ and $\alpha$. Specifically, we expand the position of the effective potential peak as a double series in the coupling parameter $\alpha$ and the inverse angular momentum parameter $\kappa^{-1}$,
\begin{equation}
    r_{\rm max}= r_{00} + r_{01}\alpha + r_{02}\alpha^2 + \ldots + \frac{r_{20} + r_{21}\alpha + \ldots}{\kappa^2} + \ldots,
    \label{eq:raprox}
\end{equation}
The coefficients $r_{ij}$ are fixed recursively: after inserting Eq.~(\ref{eq:raprox}) into $V'(r)=0$, each independent power of $\alpha$ and $\kappa^{-1}$ is set to zero. Substituting this expansion into the effective potential and using the WKB formula at the corresponding order then yields analytic estimates for the quasinormal frequencies. This approach allows us to systematically track the influence of higher-curvature corrections on QNMs at large but finite multipole numbers, thereby bridging the gap between the eikonal limit and the numerically computed low-$\ell$ spectrum.

The general analytic expressions obtained in this approach are rather cumbersome and will not be displayed in full detail here. As an illustration, we present the leading orders for scalar perturbations of the $D=5$ black hole model $(f)$, which take the form
\begin{eqnarray}\label{eq:proposalaproxd5}
\omega r_0 =
\frac{\kappa}{2} \left[1 - \frac{3\alpha}{16r_0^2} \left(1+\frac{59\alpha}{96r_0^2} \right) \right] - \frac{i(2n+1)}{2\sqrt{2}} \left[1 - \frac{7\alpha}{16r_0^2} \left(1 + \frac{59\alpha}{224r_0^2} \right) \right]
+\mathcal{O}\left(\alpha^{3},\kappa^{-1}\right).
\end{eqnarray}
We checked that all cases exhibit the expected behavior regarding the introduced corrections: for positive values of $\alpha$, both the oscillation frequencies and damping rates decrease.

\begin{figure}[ht]
    \begin{subfigure}{0.5\textwidth}
        \includegraphics[width=\linewidth]{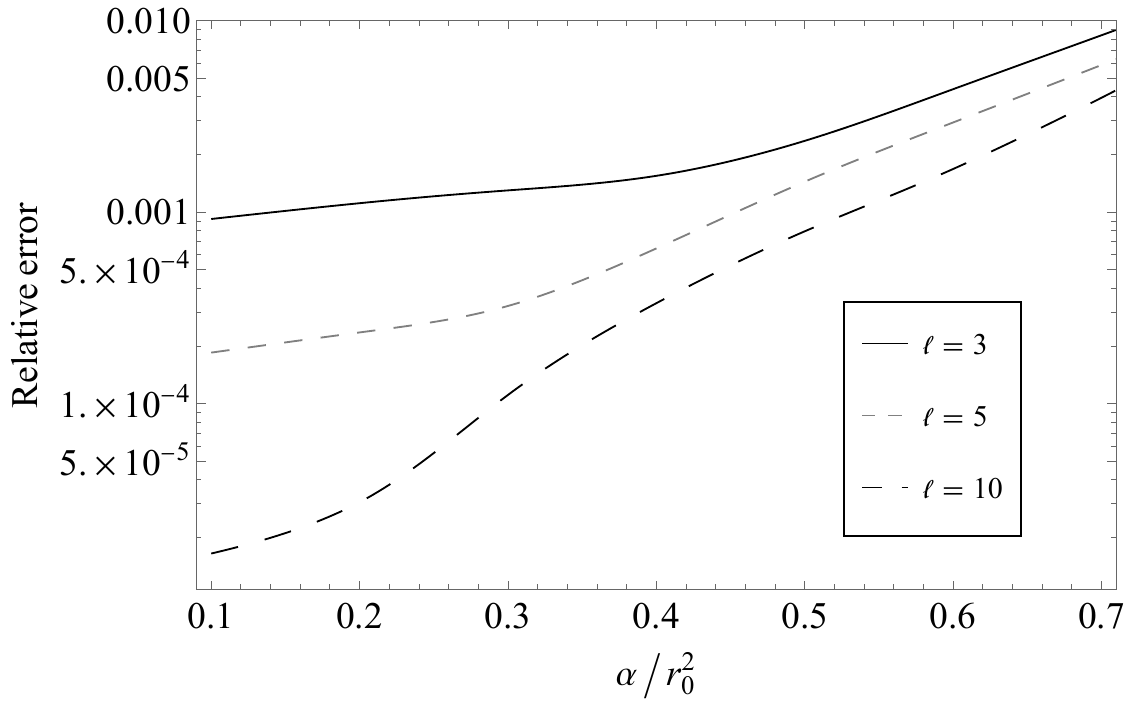}
    \end{subfigure}
    \hfill
    \begin{subfigure}{0.5\textwidth}
        \includegraphics[width=\linewidth]{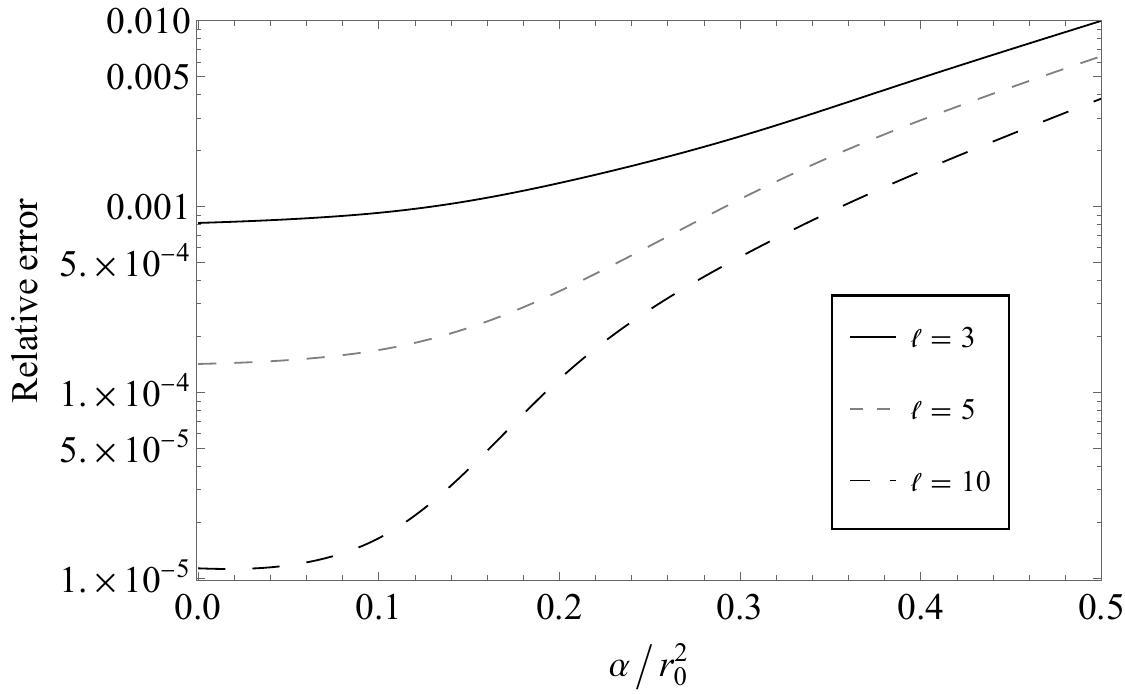}
    \end{subfigure}
    \\[2ex]
    \begin{subfigure}{0.5\textwidth}
        \includegraphics[width=\linewidth]{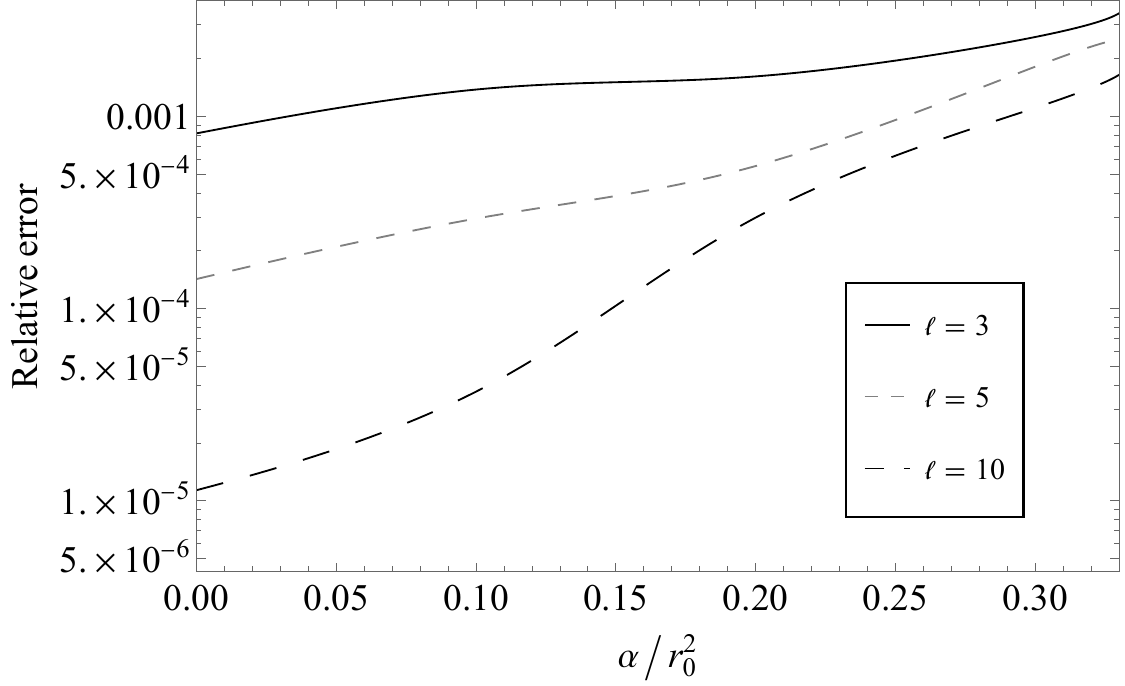}
    \end{subfigure}
    \hfill
    \begin{subfigure}{0.5\textwidth}
        \includegraphics[width=\linewidth]{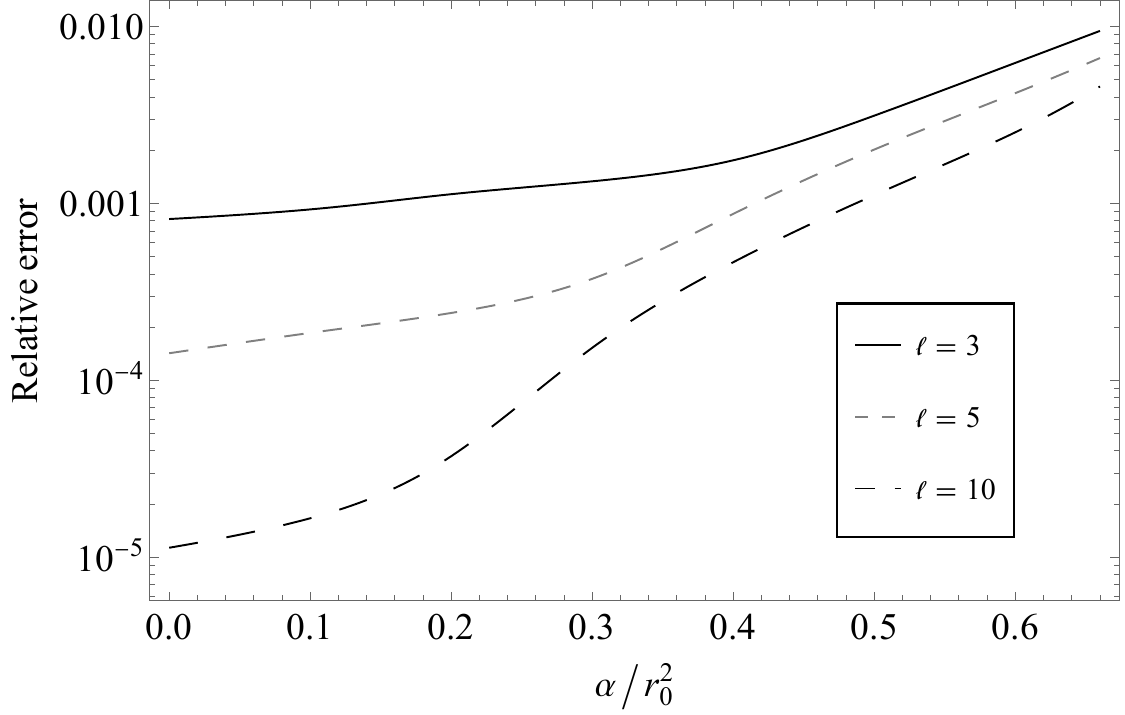}
    \end{subfigure}
    \caption{Relative error between the 13th–order WKB method with the Padé approximant ($\tilde{m}=7$) and the 3rd–order WKB formula expansion truncated at $\mathcal{O}(\alpha^{4},\kappa^{-3})$ for the scalar field with $\ell=3,5,10$ in $D=5$ regular black holes, corresponding to models (c), (d), (e), and (f), respectively (arranged from the top-left to the bottom-right panel). The vertical axis is shown on a logarithmic scale.}
    \label{fig:errorrel}
\end{figure}

Fig.~\ref{fig:errorrel} shows the relative error of the analytic approximation for the scalar field as a function of the parameter $\alpha$ for the black hole models $(c)$, $(d)$, $(e)$, and $(f)$, respectively, for different values of $\ell$ with $D=5$.

We find that, for sufficiently small values of the coupling $\alpha$, the relative error of the analytic approximation rapidly decreases as the multipole number grows. Consequently, the analytic formula correctly reproduces the asymptotic behavior of the quasinormal frequencies at large $\ell$. This confirms that the qualitative features observed in the low-multipole modes—namely, the reduction of both the oscillation frequency and the damping rate—persist for all $\ell$.

\section{Conclusions}\label{sec:conclusions}

Perturbations and quasinormal modes of black holes in theories with higher-curvature corrections have been extensively investigated in the literature, including studies of greybody factors and related phenomenological properties~\cite{Gogoi:2023ffh,Baruah:2025ifh,Lutfuoglu:2025qkt,Bolokhov:2025egl,Zinhailo:2019rwd,Konoplya:2020jgt,Konoplya:2023ppx,Blazquez-Salcedo:2020rhf,Blazquez-Salcedo:2020caw,Konoplya:2023ahd,Lutfuoglu:2025hjy,Konoplya:2025mvj,Konoplya:2017lhs,MoraisGraca:2016hmv,Chen:2015fuf}. Likewise, numerous studies have analyzed the spectra of various models of regular black holes~\cite{Bonanno:2025dry,Bronnikov:2019sbx,Bolokhov:2023ruj,Skvortsova:2024wly,Skvortsova:2024eqi,Zinhailo:2023xdz,Malik:2024qsz,Malik:2024tuf,Ge:2024vyg,Dubinsky:2024aeu,Stashko:2024wuq,Lutfuoglu:2025ohb,Skvortsova:2025cah}. In the present work, we have considered six distinct classes of theories arising from the recently formulated higher-curvature extensions of General Relativity~\cite{Bueno:2024dgm}, which, in addition to their quantum-gravity motivation, possess the important property of regularity. By combining numerical and analytic methods, we have shown that these corrections lead to a consistent qualitative modification of the quasinormal mode spectrum: both the oscillation frequencies and damping rates decrease with respect to their general relativistic counterparts. This effect is more pronounced for larger values of the higher-curvature couplings, but remains robust across all the models analyzed.

The coupling parameter $\alpha$ plays a central role in this framework. Mathematically, it controls the contribution of higher-curvature terms in the action and influences the convergence properties of the series underlying the regularization mechanism. Physically, it introduces a characteristic length scale that parametrizes deviations from General Relativity, becoming increasingly relevant in high-curvature regimes. In this sense, $\alpha$ can be associated with the scale at which the effective description of gravity breaks down, providing an estimate of the regime where new physics beyond General Relativity may emerge.

From an observational perspective, we find that the squared modulus of the quasinormal frequencies increases with the spacetime dimensionality $D$. This indicates that higher-dimensional black holes oscillate at higher frequencies while simultaneously exhibiting faster damping. Notably, the damping rate grows more rapidly than the oscillation frequency, implying that these objects behave as progressively less efficient oscillators as the number of dimensions increases~\cite{Konoplya:2011qq}. This behavior highlights the sensitivity of quasinormal modes to the dimensionality of spacetime, reinforcing their role as effective probes of the underlying gravitational theory.

Although higher-dimensional black holes are not expected to be astrophysically realized, their importance lies in their role as theoretical and phenomenological probes of gravity beyond four dimensions. In particular, in scenarios with extra dimensions, gravity may become effectively higher-dimensional at sufficiently small length scales or high energies. In such frameworks, black holes are expected to be microscopic objects whose properties encode information about the number and geometry of the extra dimensions. Consequently, their quasinormal spectra provide a potential observational window into the structure of spacetime at short distances.

In the regime where the characteristic size of the black hole is much smaller than the size of the extra dimensions, the geometry can be accurately described as $D$-dimensional. This makes higher-dimensional black holes particularly relevant as effective descriptions of trans-Planckian or near-fundamental-scale phenomena, such as those that could arise in high-energy particle collisions. It is important to stress, however, that the present analysis is performed within a semiclassical regime, where the horizon radius satisfies $r_0 \gg \ell_{\mathrm{Pl}}$. While this ensures the validity of the geometric description, a complete understanding of microscopic black holes, particularly in regimes closer to the Planck scale, would ultimately require a full theory of quantum gravity.

Our results are consistent with previous studies, particularly those involving regular black holes that do not arise from perturbative higher-curvature expansions around General Relativity \cite{Konoplya:2024kih}. The qualitative agreement between these different constructions suggests that the observed behavior of the quasinormal mode spectrum reflects general properties of regular geometries, rather than features tied to a specific model or approximation scheme. This trend also appears to be largely insensitive to the spacetime dimension and to the mechanism by which the central singularity is resolved.

Although quasinormal modes alone cannot unambiguously distinguish regular black holes from singular spacetimes, the systematic reduction in both oscillation frequencies and damping rates induced by higher-curvature corrections may serve as an indirect signature of regularity. In practice, when only a finite number of curvature terms is included, the spectrum approaches that of a fully regular configuration, rendering the two scenarios observationally indistinguishable \cite{Konoplya:2025uta}. Nevertheless, the persistence of this behavior across different models suggests that future high-precision observations could provide valuable insights into the nature of black hole interiors.

Finally, the models $(a)$–$(f)$ considered in this work should be interpreted as representative examples of higher-curvature corrections rather than definitive physical theories. From an observational standpoint, they exhibit qualitatively similar behavior, and current data do not allow one to distinguish between them. Any potential discrimination would require either significantly more precise measurements or additional theoretical input selecting a preferred structure of the higher-curvature terms.

An interesting direction for future work would be to extend the present analysis by incorporating additional physical aspects that have been explored in the literature, such as thermodynamic properties, greybody factors, and the sparsity of Hawking radiation. In particular, recent studies on black hole thermodynamics, topology, and related observable signatures suggest that combining quasinormal modes with complementary observables can provide a more complete characterization of black hole spacetimes~\cite{Alipour:2025hxa,NooriGashti:2025tde}. Since the same photon-sphere structure also controls optical observables such as the critical impact parameter, black-hole shadows, and high-frequency absorption cross sections, it would be interesting to investigate whether the regularization effects identified here produce correlated signatures in those quantities.

\ack{J. P. A. acknowledges support from the Conselho Nacional de Desenvolvimento Científico e Tecnológico (CNPq). The author is also grateful to Alexander Zhidenko for proposing the problem and reading the manuscript.}

\bibliographystyle{iopart-num}
\bibliography{bibliography}

\end{document}